\documentclass[journal]{IEEEtran}

\usepackage{pgfplots}
\usepackage[bookmarks,colorlinks]{hyperref}
\usepackage[linesnumbered,ruled,lined]{algorithm2e}
\usepackage{enumitem}
\usepackage{algpseudocode}
\usetikzlibrary{shapes.multipart,intersections}
\usepackage{cite}
\usepackage{amsmath,amssymb,amsfonts,amsthm,steinmetz}
\usepackage{graphicx}
\usepackage{mathrsfs}  
\usepackage{textcomp}
\usepackage{acronym}
\usepackage{upgreek,xspace}
\usepackage{array}
\usepackage{tikz}
\usetikzlibrary{calc}
\newcommand{\gettikzxy}[3]{%
  \tikz@scan@one@point\pgfutil@firstofone#1\relax
  \edef#2{\the\pgf@x}%
  \edef#3{\the\pgf@y}%
}
\makeatother

\usepackage[draft]{todonotes}   
\usepackage{array}      
\usepackage{multirow}   

\usepackage{esvect}

\usepackage{times}
\usepackage{bm}
\usepackage{amsmath}
\usepackage{amssymb}
\usepackage{stmaryrd}
\usepackage{babel}
\usepackage{graphics, graphicx}
\usepackage{gensymb}
\usepackage{cite}
\usepackage{enumitem}
\usepackage{url}
\newtheorem{rem}{Remark}

\begin{document}

\title{Ambiguity-Aware Segmented Estimation of \\Mutual Coupling in Large RIS: \\Algorithm and Experimental Validation}

\author{Philipp~del~Hougne,~\IEEEmembership{Member,~IEEE}
\thanks{This work was supported in part by the ANR France 2030 program (project ANR-22-PEFT-0005), the ANR PRCI program (project ANR-22-CE93-0010), the Rennes M\'etropole AES program (project ``SRI''), the European Union's European Regional Development Fund, and the French Region of Brittany and Rennes M\'etropole through the contrats de plan \'Etat-R\'egion program (projects ``SOPHIE/STIC \& Ondes'' and ``CyMoCoD'').}
\thanks{
P.~del~Hougne is with Univ Rennes, CNRS, IETR - UMR 6164, F-35000, Rennes, France (e-mail: philipp.del-hougne@univ-rennes.fr).
}
}

\maketitle

\begin{abstract}
Optimizing a real-life RIS-parametrized wireless channel with a physics-consistent multiport-network model necessitates prior remote estimation of the mutual coupling (MC) between RIS elements. The number of MC parameters grows quadratically with the number of RIS elements, posing scalability challenges. Because of inevitable ambiguities, independently estimated segments of the MC matrix cannot be easily stitched together. Here, by carefully handling the ambiguities, we achieve a separation of the full estimation problem into three sequentially treated sets of smaller problems. We partition the RIS elements into groups. \textit{First}, we estimate the MC for one group as well as the characteristics of the available loads.
\textit{Second}, we separately estimate the MC for each of the remaining groups, in each case with partial overlap with an already characterized group. \textit{Third}, we separately estimate the MC between each distinct pair of groups. Full parallelization is feasible within the second and third sets of problems, and the third set of problems can furthermore benefit from efficient initialization.
We experimentally validate our algorithm for a $4 \times 4$ MIMO channel parametrized by a 100-element RIS inside a rich-scattering environment. Our experimentally calibrated 5867-parameter multiport-network model achieves an accuracy of 40.5~dB, whereas benchmark models with limited or no MC awareness only reach 17.0~dB and 13.8~dB, respectively. Based on the experimentally calibrated models, we optimize the RIS for five wireless performance indicators. Experimental measurements with the optimized RIS configurations demonstrate only moderate benefits of MC awareness in RIS optimization in terms of the achieved performance. However, we observe that limited or no MC awareness markedly erodes the reliability of model-based predictions of the expected performance.
\end{abstract}


\section{Introduction}

Reconfigurable intelligent surfaces (RISs) enable wireless practitioners to tailor wireless channels to the needs of diverse wireless functionalities. Multiport-network theory (MNT) provides a rigorous framework to physics-consistently model such RIS-parametrized channels~\cite{gradoni_EndtoEnd_2020,matteo_universal}. Specifically, within MNT, each tunable lumped element that parametrizes a RIS element's response is treated as a lumped port terminated by a tunable load. Importantly, MNT accurately captures mutual coupling (MC) between RIS elements. To date, the main focus of theoretical MNT-based studies has been on MC-aware optimizations of the RIS configuration~\cite{gradoni_EndtoEnd_2020,qian2021mutual,abrardo2021mimo,ma2023ris,el2023optimization,li2024beyond,abrardo2024design,wijekoon2024phase,matteo_universal,nerini2024global,semmler2024performance,peng2025risnet}. However, this presupposes that the MNT model parameters are known.

Theoretical studies consider strongly simplified scenarios (e.g., minimally scattering thin-wire RIS elements in free space) for which the parameters can be determined analytically~\cite{gradoni_EndtoEnd_2020}. There are also some numerical efforts to extract the parameters for realistic RIS designs from full-wave simulations~\cite{zhang2022macromodeling,tapie2023systematic,zheng2024mutual}. However, practitioners need tools that do not require strong a priori knowledge. On the one hand, the RIS design may be completely unknown because it is proprietary, or partially unknown due to fabrication inaccuracies and component tolerances. On the other hand, the radio environment's geometry and material composition cannot be known accurately in practice unless it is free space. Techniques requiring minimal a priori knowledge to estimate the MNT model parameters given an experimental RIS-parametrized channel are thus essential. 

Experimentally, the MNT model parameters can usually \textit{not} be identified unambiguously. This limitation originates from the infeasibility of direct measurements and inevitable ambiguities in most remote measurements, as we now explain.

Directly measuring the MNT parameters would require connecting the lumped ports involved in the conceptual description of the RIS elements' tunable components to a vector network analyzer (VNA) or similar instrumentation. However, these conceptual ports associated with the RIS elements are not physically accessible in integrated RIS designs, i.e., one cannot connect them to a VNA instead of connecting them to the tunable loads.\footnote{Recent RIS prototypes with modular designs~\cite{ming2025hybrid,tapie2025beyond,del2025experimental} offer direct access to the ports associated with the RIS elements. Such modular designs are therefore valuable in early-stage academic research but integrated designs will prevail in large-scale real-life deployment because they are more compact and cheaper to manufacture at scale.} Moreover, VNAs typically have less than ten ports, whereas RISs are envisioned to have hundreds of elements. Furthermore, for most real-life applications, VNAs are prohibitively costly.
The integration of sensors into RIS elements does not appear to provide a viable alternative for directly measuring the MNT parameters. Indeed, sensor-equipped RIS elements were proposed for channel estimation assuming MC-unaware cascaded channel models~\cite{ma2020smart,alexandropoulos2023hybrid}. Directly measuring the MC between the RIS elements would require, first, synchronization of the sensors, and, second, sensors capable of emitting waves. Ultimately, the sensor network integrated into the RIS would have to offer capabilities similar to a VNA with hundreds of ports, which appears unfeasible. Altogether, the MNT model parameters thus cannot be measured directly in experiments.

Hence, experimentally, the MNT model parameters can only be estimated remotely. However, remote estimates of MNT parameters are typically ambiguous.
Indeed, an unambiguous MNT parameter estimation is only possible if each port associated with a RIS element can be terminated with three distinct and known individual loads, as well as known coupled loads~\cite{del2024virtual,del2024virtual2p0}.\footnote{Very recent work~\cite{del2025wireless} suggests that two distinct known individual loads and known coupled loads are sufficient because the latter provide effectively a third termination distinct from the two individual loads.} These conditions, under which tunable loads can serve as ``virtual'' VNA ports~\cite{del2024virtual,del2024virtual2p0}, are typically not satisfied. 
Most RIS prototypes only have two distinct individual loads (realized with a PIN diode) whose characteristics are only known approximately (due to component tolerances) or not at all. Fortunately, however, the ambiguities in the MNT model parameters are \textit{operationally irrelevant} for practitioners who use them for MC-aware RIS optimization~\cite{sol2024experimentally,del2025physics,del2025experimental}.

Few works have tackled the (inevitably ambiguous) remote MNT parameter estimation for RIS-parametrized channels without assuming strong a priori knowledge about the RIS design and/or the radio environment. The first technique, introduced and experimentally validated in~\cite{sol2024experimentally}, was a global gradient descent, seeking to simultaneously estimate all MNT parameters.\footnote{\cite{sol2024experimentally} used a coupled-dipole formulation that is operationally equivalent to the scattering-parameter representation of MNT [Sec.~II.C.2,~\cite{del2025physics}].} The same technique was applied to beyond-diagonal RIS (BD-RIS) with tunable inter-element coupling in~\cite{del2025physics} based on a physics-compliant diagonal representation of BD-RIS-parametrized channels. However, a fundamental difficulty associated with a global gradient descent is that the number of unknowns scales quadratically with the number of RIS elements. Since RISs are envisioned to consist of hundreds of RIS elements, the computational requirements for a global gradient descent can become prohibitively costly.

Recently,~\cite{del2025experimental} combined closed-form and gradient-descent steps to reduce the number of parameters estimated by gradient descent. Specifically,~\cite{del2025experimental} limited the application of gradient descent essentially to estimating the MC between the RIS elements while using singular-value decompositions (SVDs) to estimate the transmission coefficients between antennas and RIS elements and fixing the MC between the antennas based on a single reference measurement. However, the fundamental computational challenges associated with the quadratic scaling persist. 

Ideally, the estimation of the MC between the RIS elements could be separated into smaller chunks that can be independently tackled in separate (independent) gradient descents (not to be confused with alternating block gradient descent). Thereby, limitations on computational memory could be accommodated and massive parallelization could be achieved. In experiments where measurement speed is the bottleneck, one could finalize the estimation of segment parameters for which the measurements are already completed while other measurements are still ongoing.
Moreover, one may hypothesize that a gradient descent on smaller problems converges faster because the associated parameter landscape is smoother.
However, the inevitable ambiguities fundamentally thwart such a separation into independent smaller problems because independently estimated blocks of the MC matrix cannot be straightforwardly stitched together.

Given the current lack of experimental parameter estimation techniques for MNT models of RIS-parametrized channels, there is consequently also a lack of experimental MNT-based performance evaluations. While~\cite{sol2024experimentally} considered a relatively large RIS with 68 1-bit-programmable elements, the performance evaluation therein was limited to a few antenna-level metrics optimized with a dictionary search. Meanwhile,~\cite{tapie2025beyond,del2025experimental} conducted more extensive experimentally grounded performance evaluations of communications metrics for beyond-diagonal RIS and RIS with multi-bit-programmable elements, respectively, but the considered RISs were only composed of 8 RIS elements. Moreover, only~\cite{del2025experimental} conducted experimental channel measurements for the optimized RIS configurations. Altogether, the literature does not contain any reports of extensive experimental MNT-based performance evaluations for \textit{large} RIS. 

\subsection{Contributions}
In this article, we develop an ambiguity-aware segmentation of the MC parameter estimation problem for large RIS and leverage it for experimental MNT-based performance evaluations with a large-scale RIS. Our contributions are summarized as follows.
\begin{enumerate}
    \item We present a separation of the estimation of the RIS's MC parameters into three sequentially tackled sets of subproblems based on a partition of the RIS elements into small groups. The first set only comprises one subproblem. The number of subproblems in the second set scales linearly with the number of unknowns; the problems can be formulated completely independently from each other and are hence fully parallelizable. The number of subproblems in the third set scales quadratically with the number of unknowns; the subproblems are completely independent from each other and hence fully parallelizable.
    \item We identify a low-cost initialization for the third set of subproblems.
    \item We experimentally validate our technique for a 100-element RIS. Moreover, we benchmark the achieved accuracy against three simplified models. Furthermore, we systematically examine the influence of the number of antennas and the measurement noise on the achieved accuracy.
    \item We conduct MC-aware optimizations with our calibrated MNT model for five communications performance metrics. Moreover, we benchmark the achieved performances against those achieved with three simplified models.
\end{enumerate}

\subsection{Organization}
The remainder of this article is organized as follows.
In Sec.~\ref{sec_system_model}, we describe our MNT system model, as well as three low-fidelity simplified models used for benchmarking.
In Sec.~\ref{sec_param_estim_alg}, we introduce our ambiguity-aware segmented parameter estimation technique for the MNT model step-by-step. 
In Sec.~\ref{sec_exp_validation}, we describe our experimental setup and detail the experimental validation of our technique.
In Sec.~\ref{sec_exp_perf_eval}, we report extensive experimental performance evaluations based on our calibrated MNT-based model, as well as relevant benchmarking against simplified models.
In Sec.~\ref{sec_discussion}, we discuss use cases of our technique for wireless practitioners.
In Sec.~\ref{sec_conclusion}, we conclude.

\subsection{Notation}
$\mathbf{A} = \mathrm{diag}(\mathbf{a})$ denotes the diagonal matrix $\mathbf{A}$ whose diagonal entries are $\mathbf{a}$. 
$\mathbf{A}_\mathcal{BC}$ denotes the block of the matrix $\mathbf{A}$ whose row [column] indices are in the set $\mathcal{B}$ [$\mathcal{C}$]. 
$\mathcal{B}_i$ is the singleton containing the $i$th entry of $\mathcal{B}$.
$a_i$ denotes the $i$th entry of the vector $\mathbf{a}$. 
$a_{ij}$ denotes the $(i,j)$th entry of the matrix $\mathbf{A}$. 
$^\top$ and $^\dagger$ denote the transpose and transpose-conjugate operations, respectively. 
$\mathbf{1}_q$ denotes a $q\times 1$ vector whose entries are unity.

\section{System Model}
\label{sec_system_model}

In this section, we first introduce the high-fidelity physics-consistent MNT system model in Sec.~\ref{subsec_MNT}. Then, in Sec.~\ref{subsec_simplified_models}, we relate it to three simplified system models that we consider for benchmarking in this paper. Finally, in Sec.~\ref{subsec_load_characs}, we discuss the mapping from the control sequence chosen by the wireless practitioner to the physical load characteristics of the RIS elements.

\subsection{MNT System Model}
\label{subsec_MNT}

Any RIS-parametrized radio environment can be split into three entities: (i) a set of $N_\mathrm{A}$ antenna ports (via which waves can be injected and/or received), (ii) a set of $N_\mathrm{S}$ tunable lumped elements (which can be modeled as lumped ports terminated by tunable individual loads), and (iii) static scattering objects.\footnote{This statement  also applies to channels parametrized by BD-RIS~\cite{del2025physics}.} The latter comprise both environmental scattering objects as well as the structural scattering of the antennas and the RIS elements. We assume that the items captured by (iii) are linear, passive, and reciprocal; beyond these general assumptions, we do \textit{not} assume any specifics related to the radio environment or RIS element design. We further assume that all ports are sufficiently small to be described as lumped ports. Finally, we assume for simplicity that the generators and detectors that inject and receive signals via the antenna ports are matched to a common reference impedance $Z_0$. 

Irrespective of the complexity of the radio environment and the detailed RIS element design, entity (iii) is thus an $N$-port system, where $N=N_\mathrm{A}+N_\mathrm{S}$, characterized by its scattering matrix $\mathbf{S}\in\mathbb{C}^{N\times N}$ (defined using the same reference impedance $Z_0$ at all ports). Moreover, the set of $N_\mathrm{S}$ tunable loads constitutes an $N_\mathrm{S}$-port scattering system characterized by its diagonal scattering matrix $\mathbf{\Phi}\in\mathbb{C}^{N_\mathrm{S}\times N_\mathrm{S}}$. The $i$th diagonal entry of $\mathbf{\Phi}$ is the reflection coefficient $c_i$ of the $i$th individual load associated with the $i$th RIS element. We further define the vector $\mathbf{c}=[c_1, c_2, \dots, c_{N_\mathrm{S}}]\in\mathbb{C}^{N_\mathrm{S}}$ that is related to $\mathbf{\Phi}$ as follows: $\mathbf{\Phi}=\mathrm{diag}(\mathbf{c})$.
We denote by $\mathcal{T}$, $\mathcal{R}$ and $\mathcal{S}$ the sets of port indices associated with the $N_\mathrm{T}$ transmitting antennas, the $N_\mathrm{R}$ receiving antennas, and the $N_\mathrm{S}$  RIS elements, respectively, where $N_\mathrm{A}=N_\mathrm{T}+N_\mathrm{R}$.
According to MNT~\cite{matteo_universal,del2025physics}, the end-to-end wireless channel matrix $\mathbf{H}$ is defined as
\begin{equation}
    \mathbf{H} = {\mathbf{S}}_\mathcal{RT} +{\mathbf{S}}_\mathcal{RS} \left(\mathbf{\Phi}^{-1} - {\mathbf{S}}_\mathcal{SS}  \right)^{-1} {\mathbf{S}}_\mathcal{ST}.
    \label{eq1}
\end{equation}
The matrix inversion in (\ref{eq1}) can be rewritten as infinite sum whose convergence is guaranteed by passivity~\cite{del2025physics,zheng2024mutual,wijekoon2024phase}:
\begin{equation}
    \mathbf{H} = {\mathbf{S}}_\mathcal{RT} +{\mathbf{S}}_\mathcal{RS} \left[\sum_{k=0}^{\infty}(\mathbf{\Phi}{\mathbf{S}}_\mathcal{{SS}})^k\right] \mathbf{\Phi} {\mathbf{S}}_\mathcal{ST}.
    \label{eq2}
\end{equation}
The $k$th term in the infinite series corresponds to the family of paths that interact with the RIS $k$ times~\cite{del2025physics}.

\subsection{Simplified System Models}
\label{subsec_simplified_models}

Besides this high-fidelity MNT model, we consider three lower-fidelity simplified models for benchmarking.

\subsubsection{LFMNT}
Truncating the infinite sum after $k=1$ yields a low-fidelity MNT model that captures MC only to first order:
\begin{equation}
    \mathbf{H}^\mathrm{LFMNT} = {\mathbf{S}}_\mathcal{RT} +{\mathbf{S}}_\mathcal{RS} \mathbf{\Phi} {\mathbf{S}}_\mathcal{ST}+{\mathbf{S}}_\mathcal{RS} \mathbf{\Phi}{\mathbf{S}}_\mathcal{SS} \mathbf{\Phi} {\mathbf{S}}_\mathcal{ST}.
    \label{eq3}
\end{equation}

\subsubsection{CASC}
\label{subsubsec_casc}
Truncating the infinite sum after $k=0$ is equivalent to assuming that all entries of ${\mathbf{S}}_\mathcal{SS}$ vanish (which implies zero MC between RIS elements and perfect matching at each RIS element port). One recovers the widespread simplified cascaded model:
\begin{equation}
    \mathbf{H}^\mathrm{CASC} = {\mathbf{S}}_\mathcal{RT} +{\mathbf{S}}_\mathcal{RS} \mathbf{\Phi} {\mathbf{S}}_\mathcal{ST}.
    \label{eq4}
\end{equation}
This affine mapping in (\ref{eq4}) still imposes some structure inherited from (\ref{eq1}). Specifically, it still imposes that changing the configuration of a single RIS element results in a rank-one update of the channel matrix. Interestingly, a phenomenological  argument to the same effect has been described based on a ``keyhole'' intuition in~\cite{zegrar2021reconfigurable}, inspired by older seminal work on multiple-input multiple-output (MIMO) systems without RIS~\cite{chizhik2002keyholes}.

\subsubsection{LR}

A generic, unstructured, affine model is
\begin{equation}
\mathbf{H}^\mathrm{LR} = \left[ \sum_{k=1}^{N_\mathrm{S}} c_k \mathbf{A}_{k} \right] + \mathbf{B},
    \label{eq5}
\end{equation}
parametrized by $\mathbf{A}_k\in\mathbb{C}^{N_\mathrm{R}\times N_\mathrm{T}}$ and $\mathbf{B}\in\mathbb{C}^{N_\mathrm{R}\times N_\mathrm{T}}$.

\subsection{Mapping from Control Sequence to Load Characteristics}
\label{subsec_load_characs}

The wireless practitioner typically does \textit{not} know $\mathbf{\Phi}$. Moreover, the diagonal entries $c_k$ of $\mathbf{\Phi}$ can typically \textit{not} take arbitrary values only constrained by passivity. Instead, given a RIS with $m$-bit-programmable elements, the practitioner defines a RIS control sequence $\mathbf{w}\in \{1, 2, \dots, 2^m \}^{N_\mathrm{S}}$ and the resulting configurable physical RIS properties are 
\begin{equation}
    \mathbf{\Phi}=f(\mathbf{w}).
    \label{eq6}
\end{equation}
We assume that the available individual loads are the same for all RIS elements, and that their biasing is mutually independent. Thus, $f$ simply defines the $i$th diagonal entry of $\mathbf{\Phi}$ as the $w_i$th entry of the $2^m$-element vector $\mathbf{s}\in\mathbb{C}^{2^m}$ containing the realizable reflection coefficients.

\begin{rem}
In general, the realized loads at distinct RIS elements could differ by design and/or due to component tolerances, and/or they could be coupled, e.g., due to a limited power supply of the control circuitry.
\end{rem}

\section{Ambiguity-Aware Segmented \\ Parameter Estimation}
\label{sec_param_estim_alg}

In this section, we begin by precisely formulating our problem statement in Sec.~\ref{subsec_problem_statement}. Then, we provide a high-level overview of our algorithmic approach in Sec.~\ref{subsec_algorithmic_overview}.
We go on to detail our algorithmic steps in Sec.~\ref{subsec_algorithm}. Next, we provide technical details about our gradient-descent implementation in Sec.~\ref{subsec_graddesc}. Finally, we explain how we choose the parameters for our three simplified benchmark models in Sec.~\ref{subsec_benchmark_calibration}.

\subsection{Problem Statement}
\label{subsec_problem_statement}

Given the ability to measure $\mathbf{H}$ for any desired $\mathbf{w}$, we seek to estimate \textit{a} set of parameters $\tilde{\mathbf{S}}_\mathcal{RT}$, $\tilde{\mathbf{S}}_\mathcal{RS}$, $\tilde{\mathbf{S}}_\mathcal{SS}$, $\tilde{\mathbf{S}}_\mathcal{ST}$, and $\tilde{\mathbf{s}}$ such that we can use (\ref{eq6}) and then (\ref{eq1}) to map any $\mathbf{w}$ to the corresponding $\mathbf{H}$. We use the tilde to emphasize that our estimated parameters are generally different from ${\mathbf{S}}_\mathcal{RT}$, ${\mathbf{S}}_\mathcal{RS}$, ${\mathbf{S}}_\mathcal{SS}$, ${\mathbf{S}}_\mathcal{ST}$, and ${\mathbf{s}}$ due to the inevitable ambiguities.

For clarity, we reiterate that according to our system model and problem statement, \textit{first}, the reflection coefficients of the tunable loads associated with the RIS elements can only take $2^m$ distinct values but not any arbitrary value within or on the unit circle, \textit{second}, the available states of the tunable loads are the same for all RIS elements, \textit{third}, the tunable loads can be tuned independently from each other, \textit{fourth}, we do not know the characteristics of the available states of the tunable loads, \textit{fifth}, we do not have any prior knowledge about $\mathbf{S}$ other than that it is symmetric due to reciprocity and sub-unitarity due to passivity and inevitable absorption. These five points reflect the reality of our experiment described in Sec.~\ref{sec_exp_validation}.

\subsection{Algorithmic Overview}
\label{subsec_algorithmic_overview}

The crux consists in estimating the RIS's MC matrix $\tilde{\mathbf{S}}_\mathcal{SS}$. However, we also require matching estimates of $\tilde{\mathbf{S}}_\mathcal{RT}$, $\tilde{\mathbf{S}}_\mathcal{RS}$, $\tilde{\mathbf{S}}_\mathcal{ST}$, and $\tilde{\mathbf{s}}$.
Our technique consists of one preliminary phase, followed by three main phases that tackle the estimation of the RIS's MC matrix. An optional additional final phase can be used for fine-tuning.

\textit{Preliminary Phase:} We fix $\tilde{\mathbf{S}}_\mathcal{RT}$ based on a single reference measurement; moreover, based on $N_\mathrm{S}$ measurements, we fix in closed-form $\tilde{\mathbf{S}}_\mathcal{RS}^\top$ and $\tilde{\mathbf{S}}_\mathcal{ST}$ up to row-wise scaling factors. 

For the three main phases, we partition the RIS elements into groups.

\textit{Phase 1:} We jointly estimate (i) the diagonal block of the MC matrix associated with the first group of RIS elements, (ii) the associated scaling factors for the rows of $\tilde{\mathbf{S}}_\mathcal{RS}^\top$ and $\tilde{\mathbf{S}}_\mathcal{ST}$ associated with the first group of RIS elements, as well as (iii) all unknown entries of $\tilde{\mathbf{s}}$.

\textit{Phase 2:} For each remaining group of RIS elements in turn, we jointly estimate (i) the diagonal block of the MC matrix associated with that group of RIS elements, and (ii) the associated scaling factors for the rows of $\tilde{\mathbf{S}}_\mathcal{RS}^\top$ and $\tilde{\mathbf{S}}_\mathcal{ST}$ associated with that group of RIS elements. 

\textit{Phase 3:} For each pair of groups of RIS elements in turn, we estimate the corresponding off-diagonal block of the MC matrix.

\textit{Optional Final Phase:} All estimated parameters are jointly fine-tuned.

The four phases must be conducted in this order, but the subproblems within Phase 2 and Phase 3 can be fully parallelized.

\begin{figure}
\centering
\includegraphics[width=\columnwidth]{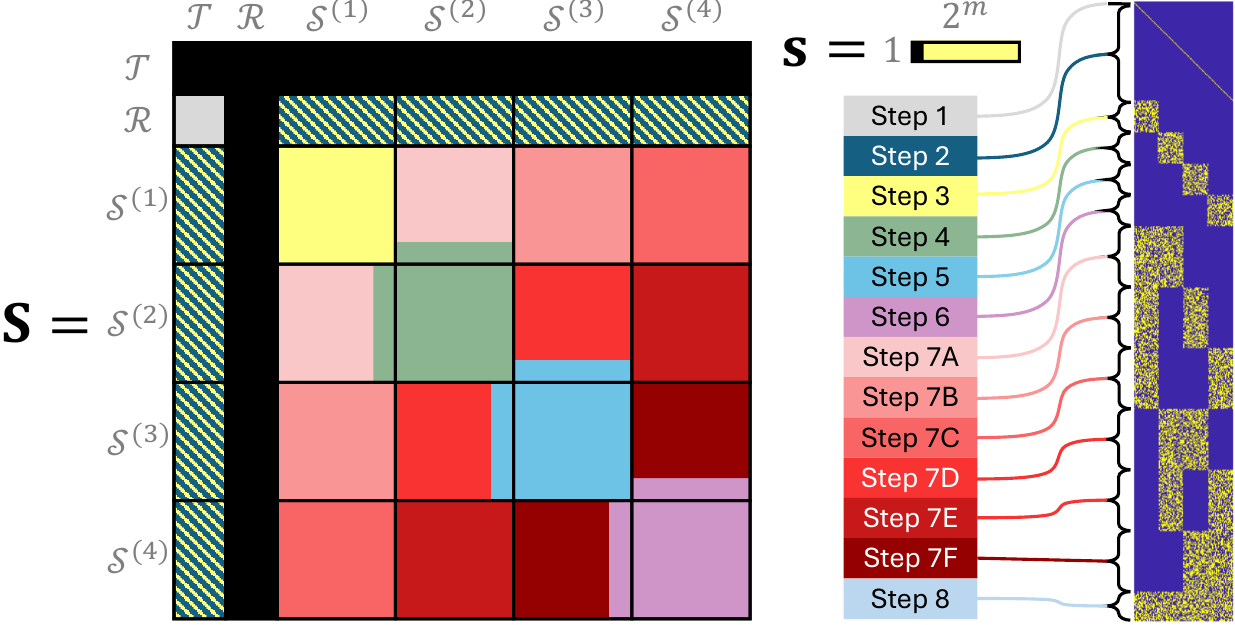}
\caption{Ambiguity-aware segmented MNT parameter estimation scheme. Colors indicate in what step a given segment is estimated. Black segments are not estimated. The estimation of hatched segments involves two steps. The right map illustrates the RIS configurations used in each step for a 1-bit programmable RIS whose reference state is colored blue. Step 8 is optional.}
\label{fig1}
\end{figure}

\subsection{Detailed Algorithmic Steps}
\label{subsec_algorithm}

\textit{Preliminary Phase}

\textit{Step 1:} We measure $\mathbf{H}(\mathbf{w}_0)$, where $\mathbf{w}_0 =\mathbf{1}_{N_\mathrm{S}}$, and define $\tilde{\mathbf{S}}_\mathcal{RT} \triangleq  \mathbf{H}(\mathbf{w}_0)$. This also implies that we fix $\tilde{s}_1=0$.

\textit{Step 2:} We measure $\mathbf{H}(\mathbf{w}_i)$, where $\mathbf{w}_i$ equals $\mathbf{w}_0$ except for the $i$th entry, and evaluate $\mathbf{\Delta}_i = \mathbf{H}(\mathbf{w}_i)-\mathbf{H}(\mathbf{w}_0)$. We compute the SVD $\mathbf{\Delta}_i = \mathbf{U}_i\mathbf{\Sigma}_i\mathbf{V}_i^\dagger$ and identify the first left singular vector $\mathbf{u}_i$ and the first right singular vector $\mathbf{v}_i$. We define $\tilde{\mathbf{S}}_{\mathcal{RS}_i} \triangleq x_i \mathbf{u}_i$ and $\tilde{\mathbf{S}}_{\mathcal{S}_i\mathcal{T}} \triangleq y_i \mathbf{v}_i^\dagger$, where $x_i$ and $y_i$ are scaling factors that remain to be determined. 
We repeat this procedure for each RIS element in turn.

These first two steps are identical to~\cite{del2025experimental}. MC does not appear to play an explicit role thus far because the rank-one property of updating a single RIS element's configuration is also found in the MC-unaware CASC model, as mentioned earlier in Sec.~\ref{subsubsec_casc}. 
However, it is the presence of MC that prevents us from using multi-flip updates with respect to $\mathbf{w}_0$ as done in the MC-unaware method in~\cite{zegrar2021reconfigurable}. Indeed, the method in~\cite{zegrar2021reconfigurable} cannot replace our Step 2 because in our problem statement MC is not negligible, $c_k$ cannot take any arbitrary value within or on the unit circle, and $\mathbf{s}$ is not known.

\textit{Phase 1}

Now, we deviate from~\cite{del2025experimental} because we segment the estimation of $\tilde{\mathbf{S}}_\mathcal{SS}$. We partition the $N_\mathrm{S}$ RIS elements into $g$ non-overlapping groups of $e$ RIS elements (except for the last group which contains $\lfloor \frac{N_\mathrm{S}}{g} \rfloor$ elements); the set of port indices associated with the $i$th group is $\mathcal{S}^{(i)}$, such that $\bigcup_{i=1}^{g} \mathcal{S}^{(i)} = \mathcal{S}$. (See also Remark~\ref{Remark4}.)

We introduce $\mathbf{x} = [x_1, x_2, \dots, x_{N_\mathrm{S}}]\in\mathbb{C}^{N_\mathrm{S}}$ and $\mathbf{y} = [y_1, y_2, \dots, y_{N_\mathrm{S}}]\in\mathbb{C}^{N_\mathrm{S}}$ for notational ease.

\textit{Step 3:} We measure $\mathbf{H}(\mathbf{w})$ for $n_1$ known realizations of $\mathbf{w}$ which equal $\mathbf{w}_0$ except for the entries associated with $\mathcal{S}^{(1)}$ which are chosen randomly. Using gradient descent, we jointly estimate $\tilde{\mathbf{S}}_{\mathcal{S}^{(1)}\mathcal{S}^{(1)}}$, $\tilde{\mathbf{s}}$ (except for its first entry that we fixed to 0 in Step 1), $\mathbf{x}_{\mathcal{S}^{(1)}}$, and $\mathbf{y}_{\mathcal{S}^{(1)}}$.

As illustrated in Fig.~\ref{fig1}, this Step 3 identifies the top left diagonal block of $\tilde{\mathbf{S}}_\mathcal{SS}$, and it simultaneously fixes the scaling factors on the associated rows and columns of $\tilde{\mathbf{S}}_\mathcal{ST}$ and $\tilde{\mathbf{S}}_\mathcal{RS}$, respectively, as well as the entries of $\tilde{\mathbf{s}}$ (except for the first one). It thus resembles Step 3 in~\cite{del2025experimental} except that it is limited to $\mathcal{S}^{(1)}$ here instead of being applied globally to $\mathcal{S}$ as in~\cite{del2025experimental}. 

\textit{Phase 2}

\textit{Step 4:} We now seek to estimate $\tilde{\mathbf{S}}_{\mathcal{S}^{(2)}\mathcal{S}^{(2)}}$, and to simultaneously fix  $\mathbf{x}_{\mathcal{S}^{(2)}}$, and $\mathbf{y}_{\mathcal{S}^{(2)}}$. We proceed similarly to Step 3 except for two modifications that are required to ensure compatibility with the ambiguous choices made in Step 3. On the one hand, we maintain $\tilde{\mathbf{s}}$ fixed after Step 3. On the other hand, instead of using random configurations of the RIS elements in $\mathcal{S}^{(2)}$ alone, we seek some overlap with the RIS elements in  $\mathcal{S}^{(1)}$. To that end, we define $\mathcal{X}^{(2)} = \mathcal{S}^{(2)} \cup \mathcal{V}^{(1)}$, where $\mathcal{V}^{(1)}$ is a set containing $e\gg v\geq 1$ distinct entries of $\mathcal{S}^{(1)}$ (see also Remark~\ref{Remark5}). 
We measure $\mathbf{H}(\mathbf{w})$ for $n_2$ known realizations of $\mathbf{w}$ which equal $\mathbf{w}_0$ except for the entries associated with $\mathcal{X}^{(2)}$ which are chosen randomly. Using gradient descent, we jointly estimate $\tilde{\mathbf{S}}_{\mathcal{S}^{(2)}\mathcal{S}^{(2)}}$, $\tilde{\mathbf{S}}_{\mathcal{S}^{(1)}\mathcal{S}^{(2)}}=\tilde{\mathbf{S}}_{\mathcal{S}^{(2)}\mathcal{S}^{(1)}}^\top$, $\mathbf{x}_{\mathcal{S}^{(2)}}$, and $\mathbf{y}_{\mathcal{S}^{(2)}}$.

\textit{Step ($4+i$), where $1\leq i < g$:} We repeat Step 4 for the $(i+1)$th diagonal block of $\tilde{\mathbf{S}}_\mathcal{SS}$ by using an overlap with the $v$ last RIS elements from $\mathcal{S}^{(i)}$. We jointly estimate $\tilde{\mathbf{S}}_{\mathcal{S}^{(i+1)}\mathcal{S}^{(i+1)}}$, $\tilde{\mathbf{S}}_{\mathcal{S}^{(i)}\mathcal{S}^{(i+1)}}=\tilde{\mathbf{S}}_{\mathcal{S}^{(i+1)}\mathcal{S}^{(i)}}^\top$, $\mathbf{x}_{\mathcal{S}^{(i+1)}}$, and $\mathbf{y}_{\mathcal{S}^{(i+1)}}$. 
Alternatively, Step ($4+i$) can be conducting using an overlap with the $v$ last RIS elements from $\mathcal{S}^{(1)}$ instead of the $v$ last RIS elements from $\mathcal{S}^{(i)}$. This alternative is appealing to maximize computational parallelization because it allows one to independently conduct the computational efforts of the $g-1$ steps that determine all but the first diagonal blocks of $\tilde{\mathbf{S}}_\mathcal{SS}$. 

\textit{Phase 3}

\textit{Step ($4+g+i$), where $1\leq i \leq h$ and $h=\frac{g(g-1)}{2}$:} We now seek to estimate the $i$th off-diagonal block of $\tilde{\mathbf{S}}_\mathcal{SS}$ associated with the groups of RIS elements identified by $\mathcal{S}^{(k)}$ and $\mathcal{S}^{(l)}$, where $k\neq l$. 
We measure $\mathbf{H}(\mathbf{w})$ for $n_3$ known realizations of $\mathbf{w}$ which equal $\mathbf{w}_0$ except for the entries associated with $\mathcal{U}^{(kl)} = \mathcal{S}^{(k)} \cup \mathcal{S}^{(l)}$ which are chosen randomly. Using gradient descent, we estimate $\tilde{\mathbf{S}}_{\mathcal{S}^{(k)}\mathcal{S}^{(l)}} = \tilde{\mathbf{S}}_{\mathcal{S}^{(l)}\mathcal{S}^{(k)}}^\top$. If some entries of $\tilde{\mathbf{S}}_{\mathcal{S}^{(k)}\mathcal{S}^{(l)}}$ were already fixed in a previous step as part of the overlapping, we do not re-estimate them. Instead of initializing $\tilde{\mathbf{S}}_{\mathcal{S}^{(k)}\mathcal{S}^{(l)}}$ randomly for the gradient descent, we use a low-fidelity estimate obtained via multi-variable linear regression by assuming a truncation of (\ref{eq2}) after $k=1$. The resulting expression 
\begin{equation}
\begin{split}
    \mathbf{H} \approx \ &\tilde{\mathbf{S}}_\mathcal{RT} +\tilde{\mathbf{S}}_{\mathcal{R}\mathcal{U}^{(kl)}} \tilde{\mathbf{\Phi}}_{\mathcal{U}^{(kl)}\mathcal{U}^{(kl)}} \tilde{\mathbf{S}}_{\mathcal{U}^{(kl)}\mathcal{T}}\\+\ &\tilde{\mathbf{S}}_{\mathcal{R}\mathcal{U}^{(kl)}} \tilde{\mathbf{\Phi}}_{\mathcal{U}^{(kl)}\mathcal{U}^{(kl)}}\tilde{\mathbf{S}}_{\mathcal{U}^{(kl)}\mathcal{U}^{(kl)}} \tilde{\mathbf{\Phi}}_{\mathcal{U}^{(kl)}\mathcal{U}^{(kl)}} \tilde{\mathbf{S}}_{\mathcal{U}^{(kl)}\mathcal{T}}    
\end{split}
\label{eq7}
\end{equation}
is analogous to (\ref{eq3}), and clearly affine in the sought-after $\tilde{\mathbf{S}}_{\mathcal{U}^{(kl)}\mathcal{U}^{(kl)}}$ of which  $\tilde{\mathbf{S}}_{\mathcal{S}^{(k)}\mathcal{S}^{(l)}}$ is a block. Because $\mathbf{H}$ is known from measurements and all other variables in (\ref{eq7}) except for $\tilde{\mathbf{S}}_{\mathcal{S}^{(k)}\mathcal{S}^{(l)}}$ were fixed in previous steps, (\ref{eq7}) can indeed be solved for $\tilde{\mathbf{S}}_{\mathcal{S}^{(k)}\mathcal{S}^{(l)}}$ via multi-variable linear regression given the $n_3$ matching pairs of $\mathbf{H}$ and $\tilde{\mathbf{\Phi}}$. Due to the approximate nature of (\ref{eq7}), this yields only a low-fidelity estimate of $\tilde{\mathbf{S}}_{\mathcal{S}^{(k)}\mathcal{S}^{(l)}}$. However, this is nonetheless useful to initialize the gradient descent that seeks the high-fidelity estimate of $\tilde{\mathbf{S}}_{\mathcal{S}^{(k)}\mathcal{S}^{(l)}}$. Because these $h$ steps aimed at retrieving off-diagonal blocks of $\tilde{\mathbf{S}}_\mathcal{SS}$ are mutually independent, they can be fully parallelized.

\textit{Optional Final Phase}

\textit{Final Step (optional):} We measure $\mathbf{H}(\mathbf{w})$ for $n_4$ known, random realizations of $\mathbf{w}$. Using gradient descent, we jointly refine all previous estimates of $\tilde{\mathbf{S}}_\mathcal{RT}$, $\tilde{\mathbf{S}}_\mathcal{RS}$, $\tilde{\mathbf{S}}_\mathcal{SS}$, $\tilde{\mathbf{S}}_\mathcal{ST}$, and $\tilde{\mathbf{s}}$, which serve as initialization. If this optional final step for global fine-tuning is used, we denote the model with ``MNTGFT''; otherwise, we denote the model with ``MNT''.

The total number of measurements is
\begin{subequations}
\begin{equation}
    n_\mathrm{MNT} = 1+N_\mathrm{S}+ n_1 + (g-1) n_2 + \frac{g(g-1)}{2} n_3.
\end{equation}
\begin{equation}
    n_\mathrm{MNTGFT} = 1+N_\mathrm{S}+ n_1 + (g-1) n_2 + \frac{g(g-1)}{2} n_3 + n_4.
\end{equation}
\end{subequations}

\begin{rem}
An affine initialization akin to the one used in Phase 3 to estimate the off-diagonal blocks of $\tilde{\mathbf{S}}_\mathcal{SS}$ is \textit{not} possible for any of the steps in Phase 1 or Phase 2 that estimate the diagonal blocks of $\tilde{\mathbf{S}}_\mathcal{SS}$. The reason is that the steps in Phase 2 simultaneously fix the scaling factors associated with the corresponding blocks of $\tilde{\mathbf{S}}_\mathcal{ST}$ and $\tilde{\mathbf{S}}_\mathcal{RS}$. The step in Phase 1 identifying the first diagonal block moreover fixes all but the first of the entries of $\tilde{\mathbf{s}}$ that map into $\tilde{\mathbf{\Phi}}$ via  (\ref{eq6}). Thus, the diagonal blocks are not the only unknowns in the steps that determine them. This prevents us from obtaining a low-fidelity estimate of a diagonal block by solving an approximate model originating from a truncation of (\ref{eq2}) at $k=0$, akin to (\ref{eq4}), via multi-variable linear regression.
\end{rem}

\begin{rem}
For the sufficiently constrained, unambiguous parameter estimation underlying the related but distinct problem statement associated with the Virtual VNA technique, a purely closed-form version exists~\cite{del2024virtual,del2024virtual2p0}. Whether a similar purely closed-form version can be identified  for our inevitably ambiguous problem statement in Sec.~\ref{subsec_problem_statement} is beyond the scope of this paper. We deliberately opt for gradient descent here because it has proven to be more noise-robust in the Virtual VNA context~\cite{tapie2025scalable}. However, it seems clear that without any a priori knowledge of $\mathbf{s}$, one requires gradient descent at least in Step 4 to identify $\tilde{\mathbf{s}}$; the case of $m=1$ (1-bit-programmable RIS elements) may be an exception.
\end{rem}

\subsection{Gradient Descent Implementation}
\label{subsec_graddesc}

All gradient descents are implemented in TensorFlow. In each case, the cost $\mathcal{C}$ to be minimized is defined as
\begin{equation}
    \mathcal{C}
= \left\langle \frac{\displaystyle\sum_{i,j} \bigl|\,h^\mathrm{PRED}_{ij} - h^\mathrm{MEAS}_{ij}\bigr|}
       {\displaystyle\sum_{i,j} \bigl|\,h^\mathrm{MEAS}_{ij}\bigr|} \right\rangle,
\end{equation}
where the superscripts PRED and MEAS indicate whether the channel gains are measured or predicted using (\ref{eq1}), and $\langle \cdot \rangle$ indicates averaging over a batch of RIS configurations. Randomly initialized variables are drawn from a truncated normal
distribution (mean: 0, standard deviation: 0.1). The batch size is 300 unless fewer examples are available. We use the Adam method for
stochastic optimization with an initial step size of $10^{-3}$ that is gradually reduced by up to two orders of magnitude as a function of $\mathcal{C}$. We stop training after 4000 iterations and restore the parameter set that yielded the lowest seen cost.

\subsection{Benchmark Models}
\label{subsec_benchmark_calibration}

We determine the parameters required by the three benchmark models introduced in Sec.~\ref{subsec_simplified_models} as follows.

\subsubsection{LFMNT}
We proceed as for the full MNT model except that skip the gradient descents in Phase 3 that estimates off-diagonal blocks of the MC matrix. Instead, we use the initialization obtained via multi-variable linear regression as the final value.

\subsubsection{CASC}
We use the parameters estimated for the full MNT model except that we set all entries of $\tilde{\mathbf{S}}_\mathcal{SS}$ to zero.

\subsubsection{LR}
We estimate the entries of $\mathbf{A}_k$ and $\mathbf{B}$ via multi-variable linear regression using the $n_4$ matching pairs of $\mathbf{\Phi}$ and $\mathbf{H}$ measured in Step 8.

\section{Experimental Validation}
\label{sec_exp_validation}

In this section, we describe in Sec.~\ref{subsec_exp_setup} our experimental setup; then, we report in Sec.~\ref{subsec_exp_param_estim} the experimental validation and analysis of our parameter estimation technique described in Sec.~\ref{sec_param_estim_alg}.

\subsection{Experimental Setup}
\label{subsec_exp_setup}

\begin{figure}
\centering
\includegraphics[width=0.7\columnwidth]{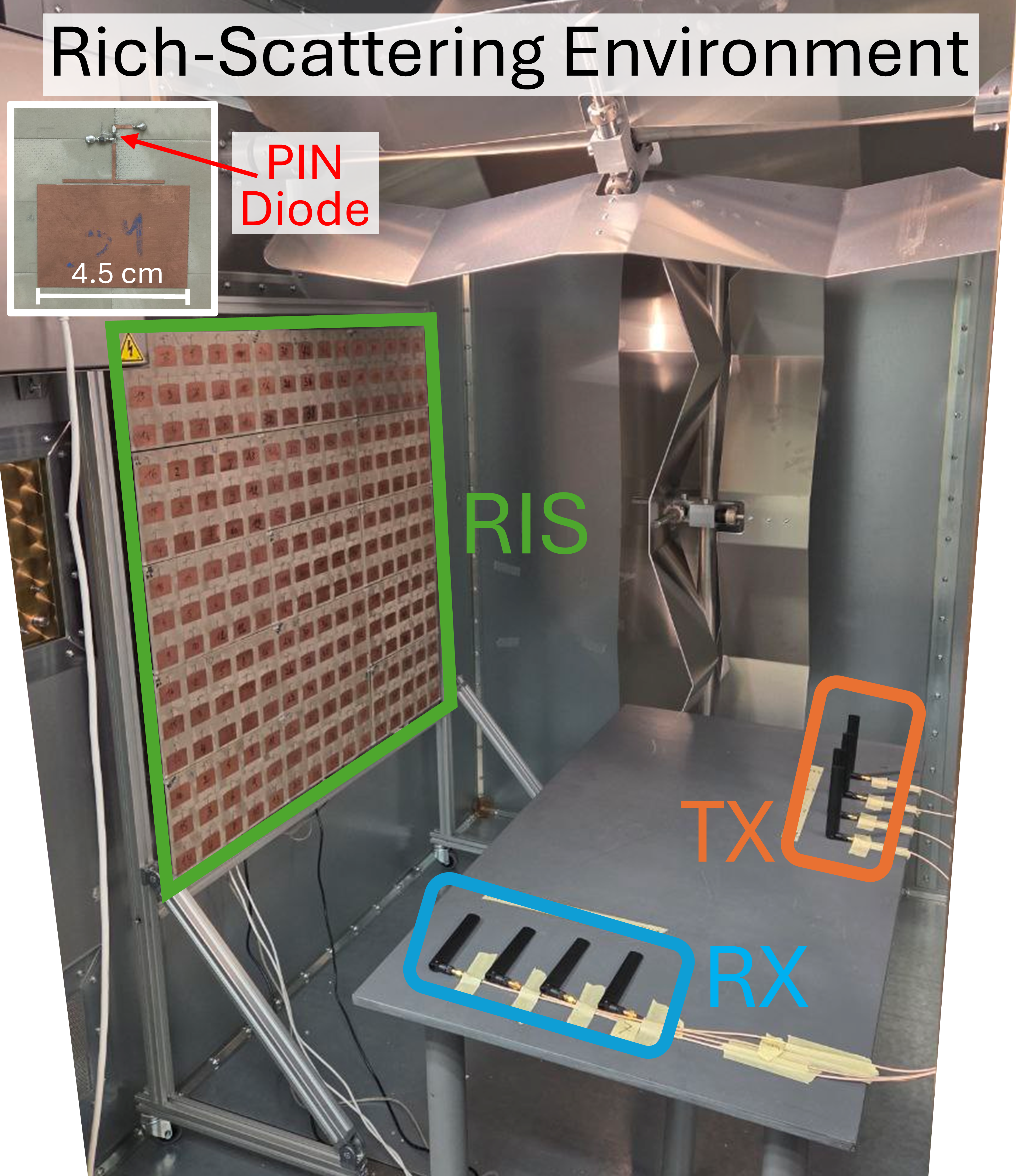}
\caption{Experimental setup (inset shows RIS element).}
\label{fig2}
\end{figure}

Our RIS prototype is displayed in Fig.~\ref{fig2} and comprises 225 1-bit programmable RIS elements of which we use 100 in this paper; the remaining ones are kept in their reference state throughout all experiments. Specifically, each RIS element is a half-wavelength patch equipped with a PIN diode whose bias voltage we control individually. The inter-element spacing is roughly half a wavelength. The RIS elements' reflection coefficients under normal incidence associated with the two possible states roughly have a phase difference of $\pi$ at 2.45~GHz. RIS design details are discussed in~\cite{ahmed2025over}. 

The pivotal technical detail of our RIS design that is relevant to the applicability of the MNT model described in Sec.~\ref{sec_system_model} is that the size of the PIN diode is tiny compared to the wavelength (see inset in Fig.~\ref{fig2}). Indeed, it is the PIN diode that we describe as lumped port terminated by a tunable load. Meanwhile, the size of the RIS element being on the order of half a wavelength and thus not electrically small is irrelevant for the applicability of the MNT model. Indeed, the RIS element void of the PIN diode constitutes a static scatterer that is treated as part of the radio environment. A direct consequence of our system model is that $\mathbf{\Delta}_i$ from Step 2 is a rank-one matrix~\cite{prod2023efficient,sol2024optimal,del2024virtual}. Experimentally, we have verified that this is approximately true by confirming that the ratio between the first and second singular value of $\mathbf{\Delta}_i$ is large (greater than ten).

Our transmit and receive arrays each comprise four parallel, half-wavelength-spaced antennas (ANT-W63WS2-SMA). 
As seen in Fig.~\ref{fig2}, they are placed alongside our RIS inside a reverberation chamber (1.75~$\times$~1.5~$\times$~2~$\mathrm{m}^3$) which constitutes an unknown, static, rich-scattering environment. To minimize the direct line-of-sight channel, the transmitting and receiving antennas are oriented perpendicular to each other.
We use an eight-port VNA (two cascaded Keysight P5024B 4-port VNAs) to measure $\mathbf{H}$ at our operating frequency of 2.45~GHz. 

We conduct our experiment under well-controlled conditions: The signal-to-noise ratio (SNR) is 55.3~dB (estimated based on repeated measurements in quick succession of $\mathbf{H}$ for the same $\mathbf{w}$) and the stability is 45.8~dB (estimated like the SNR, but based on measurements taken intermittently over the course of our experiment).

We emphasize that we dispose only of minimal a priori knowledge about our experimental system, namely that all components are reciprocal and passive, and that all tunable loads are identical and have two possible states. We do not assume to know the two states' characteristics, nor the RIS element design, nor any specifics about our rich-scattering radio environment.

\subsection{Experimental MNT Model Parameter Estimation}
\label{subsec_exp_param_estim}

We now experimentally validate the algorithm described in Sec.~\ref{sec_param_estim_alg} for our experimental $4\times 4$ MIMO channel parametrized by our RIS prototype with 100 1-bit-programmable elements, i.e., with $N_\mathrm{R}=N_\mathrm{T}=4$, $N_\mathrm{S}=100$, and $m=1$. 
The number of unknown, complex-valued, scalar MNT model parameters to be estimated is
\begin{equation}
    n_\mathrm{u} = N_\mathrm{R}N_\mathrm{T}+(N_\mathrm{R}+N_\mathrm{T})N_\mathrm{S} + \frac{N_\mathrm{S}(N_\mathrm{S}+1)}{2}+(2^m-1), 
    \label{eq10}
\end{equation}
which evaluates to $n_\mathrm{u}=5867$ for our problem.

We partition the $N_\mathrm{S}=100$ RIS elements into $g=4$ groups of $e=25$ elements. 
We conduct the required measurements following the description in Sec.~\ref{sec_param_estim_alg} with $n_1=n_2=\frac{e(e+1)}{2}=325$, $n_3 = e^2=625$, $n_4 = 300$, and $v=4$ (see Remark~\ref{Remark4} and Remark~\ref{Remark5} below). Because the measurement speed is the bottleneck in our experiments, we cannot exploit the full parallelizability within Phase 2 and Phase 3. Instead, we conduct the computations for any given subproblem as soon as the required data is available, while measurements for other subproblems may still be ongoing. For Phase 2, we choose a sequential overlap as illustrated in Fig.~\ref{fig1}. The values of $n_1$, $n_2$, and $n_3$ are chosen generously so that we can test our algorithm under ideal conditions without being limited by the available data; we explore the influence of limited data by only using a fraction $p$ of the available measurements for each segment. The larger value $n_3$ (compared to $n_1=n_2$) reflects the larger number of unknowns in off-diagonal blocks due to reciprocity. 
We furthermore measure $\mathbf{H}$ for $n_5=30$ previously unseen random RIS configurations to assess the accuracy of our models.

\begin{rem}
\label{Remark4}
The choice of $g$ should first and foremost be informed by the available computational resources to ensure that each subproblem's complexity is manageable. Secondary considerations relate to trading off potential improvements in noise robustness for larger values of $g$ (stronger variations of $\mathbf{H}$ are easier to measure in the presence of noise) and potential improvements in the smoothness of the parameter landscapes when the subproblems are smaller. Systematic investigations of these secondary considerations are beyond the scope of this paper. In our experiment with $N_\mathrm{S}=100$, $g=4$ is a reasonable choice to validate our technique.
\end{rem}

\begin{rem}
\label{Remark5}
$v$ should be significantly smaller than $e$ or else the segmentation is effectively coarser than suggested by the value of $g$. At the same time, $v$ should be sufficiently large to allow for a robust alignment of the ambiguities. While $v=1$ is sufficient under ideal conditions, it may be insufficient if the relevant RIS element happens to only weakly influence the channel or be defective. In our experiment with $e=25$, $v=4$ is a reasonable choice.
\end{rem}

To quantify the model accuracy, we use the following metric that is defined similar to an SNR, treating the model error as ``noise'':
\begin{equation}
    \zeta = \frac{\mathrm{SD}\left[h_{ij}^\mathrm{MEAS}(\mathbf{w})\right]}{\mathrm{SD}\left[h_{ij}^\mathrm{MEAS}(\mathbf{w}) - h_{ij}^\mathrm{PRED}(\mathbf{w})\right]},
    \label{eq_zeta}
\end{equation}
where $\mathrm{SD}$ denotes the standard deviation across all entries of $\mathbf{H}$ and the $n_5=30$ random, unseen realizations of $\mathbf{w}$; the superscripts denote the experimentally measured ground truth (MEAS) and the model's prediction (PRED). 

\begin{figure}
\centering
\includegraphics[width=\columnwidth]{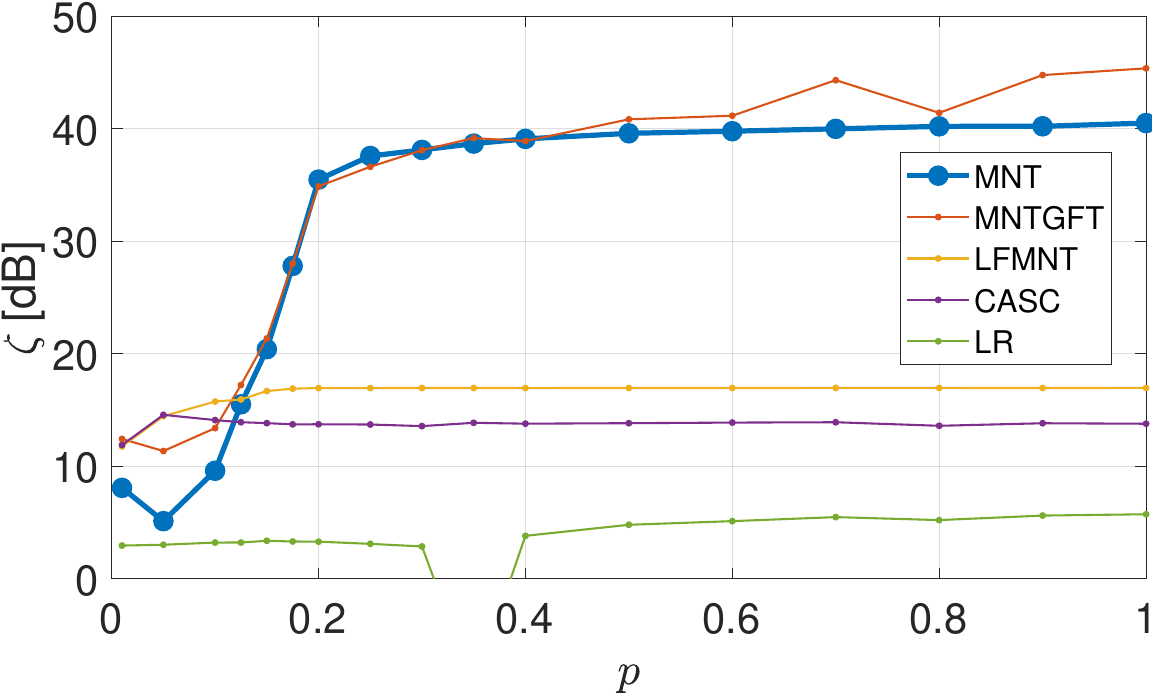}
\caption{Model accuracy metric $\zeta$ defined in (\ref{eq_zeta}) as a function of the fraction $p$ of the available data used to estimate the model parameters, for the five considered cases. }
\label{fig3}
\end{figure}

With sufficient measurements, our technique from Sec.~\ref{sec_param_estim_alg} achieves an accuracy of 40.5~dB (see MNT curve for $p=1$ in Fig.~\ref{fig3}). The amount of available data can be reduced to $p=0.4$ without any notable effect on the accuracy, and even to $p=0.2$ without any strong effect on the accuracy. Below $p=0.2$, the accuracy drops rapidly as $p$ is reduced further. It hence appears that the parameter estimation problem is ill-posed for $p<0.2$. At $p=0.2$, the total number of measured complex-valued scalars is $N_\mathrm{R} N_\mathrm{T} n_\mathrm{MNT} = 16483$, which is roughly 2.8 times the number of unknown complex-valued scalar MNT parameters. We hypothesize that this factor of 2.8 would drop close to unity only in the absence of all experimental and numerical noise. Due to the non-linear nature of our system model in (\ref{eq1}) the relatively low measurement noise and instabilities in our experiment are seemingly not negligible.

We see in Fig.~\ref{fig3} that the MNT accuracy drops monotonically as $p$ is decreased (except for the regime of $p<0.1$ in which the problem is severely ill-posed). The same is not true for MNTGFT which does \textit{not} offer consistent gains compared to MNT. 
Meanwhile, the LFMNT model, aware of MC only to first order, only achieves accuracy up to 16.9~dB. The MC-unaware CASC model only achieves accuracies up to 13.8~dB. These results highlight the importance of capturing MC effects with high fidelity for accurate modelling. The completely unstructured LR model's accuracy does not exceed 5.7~dB.

\begin{figure}
\centering
\includegraphics[width=\columnwidth]{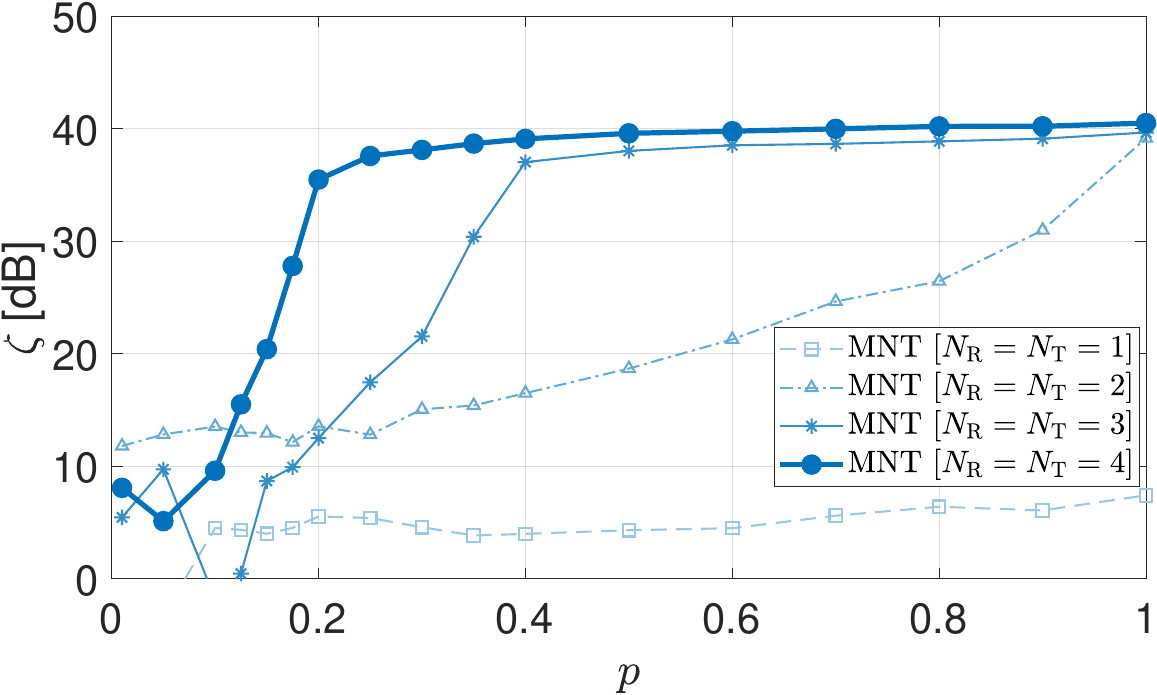}
\caption{MNT model accuracy as a function of the fraction $p$ of the available data used to estimate the model parameters, for four differently dimensioned MIMO scenarios (MIMO collapses to SISO for $N_\mathrm{R}=N_\mathrm{T}=1$). }
\label{fig4}
\end{figure}

Next, we investigate the influence of the dimensions of the MIMO system on the achieved accuracy. By ignoring parts of the measured $4 \times 4$ MIMO channel, we can consider scenarios with a $3\times 3$ channel, a $2\times 2$ channel, and a $1 \times 1$ channel (which is a single-input single-output (SISO) scenario). 
For smaller MIMO systems, $n_\mathrm{u}$ is slightly lower, as seen in (\ref{eq10}); however, $n_\mathrm{u}$ is dominated by the fixed size of the $\tilde{\mathbf{S}}_\mathcal{SS}$ block for large RIS like ours. At the same time, because each measurement yields $N_\mathrm{R}N_\mathrm{T}$ independent complex-valued scalars, the number of independent pieces of information acquired per measurement drops quadratically with the MIMO system dimension. The latter effect dominates the former, which is why we observe in Fig.~\ref{fig4} that more measurements are required for smaller MIMO systems to estimate the MNT parameters with comparable accuracy. Specifically, the $3\times 3$ MIMO system reaches a comparable accuracy of 37.0~dB only at $p=0.4$ (with marginal benefits of even larger $p$, reaching an accuracy of 39.7~dB at $p=1$), and the $2\times 2$ MIMO system reaches a comparable accuracy of 39.2~dB only at $p=1$. For the SISO system, the parameter estimation appears to be always ill-posed for the range of considered values of $p$, and the accuracy never even reaches 10~dB. The ratio of the number of required measurements to $n_\mathrm{u}$ increases as the MIMO system dimension decreases, from 2.8 for $4\times 4$, via 3.3 for $3\times 3$, to 3.8 for $2\times 2$. We hypothesize that there is no easy  way of predicting the minimal number of required measurements due to the complicated interplay between noise and the system model's non-linearity.

\begin{figure}
\centering
\includegraphics[width=\columnwidth]{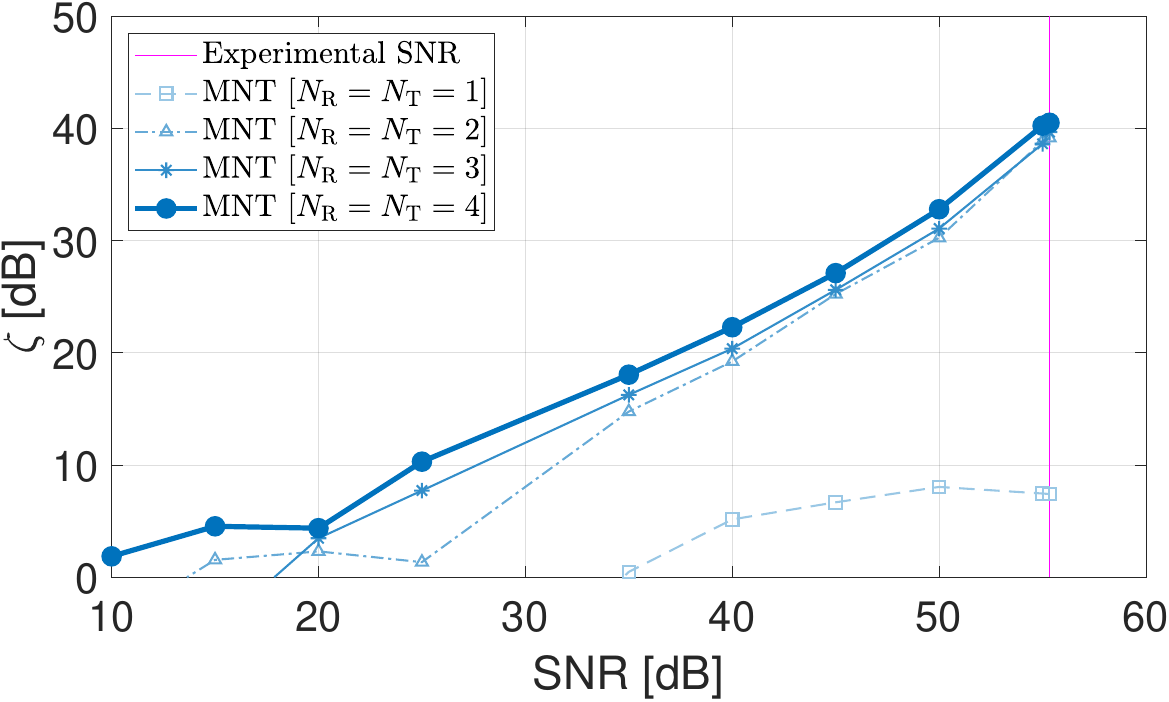}
\caption{MNT model accuracy as a function of the measurement SNR, for four differently dimensioned MIMO scenarios (MIMO collapses to SISO for $N_\mathrm{R}=N_\mathrm{T}=1$) with $p=1$. }
\label{fig5}
\end{figure}

Finally, we now examine the influence of measurement noise on the accuracy. We consider for simplicity $N_\mathrm{R}=N_\mathrm{T}=4$ and $p=1$. To increase the noise level, we synthetically add complex-valued Gaussian noise to the measured channel matrices before applying our technique from Sec.~\ref{sec_param_estim_alg}. The largest considered value of SNR is the one imposed by our experimental measurement noise (i.e., without adding any synthetic noise). We expect our algorithm to be generally quite sensitive to noise because it essentially extracts the MNT model parameters from (minute) changes of the MIMO channel as a function of the RIS configuration. 
We observe indeed in Fig.~\ref{fig5} that our accuracy metric $\zeta$~[dB] scales roughly linearly with the SNR~[dB]. The trend is very similar for all three considered MIMO systems, which reached similarly high accuracies at $p=1$ in Fig.~\ref{fig4}.
In principle, the use of more measurements (i.e., larger values of $p$) can flexibly counteract the detrimental effect of measurement noise; to what extent such an approach is practical will depend on use-case specifics.

\section{Experimental MNT-Based \\Performance Evaluation}
\label{sec_exp_perf_eval}

We have observed substantial accuracy differences between the considered, experimentally calibrated models in Sec.~\ref{subsec_exp_param_estim}. However, it remains unclear to what extent wireless practitioners require the high MNT model accuracy. To answer this question, we perform extensive experimental performance evaluations in this section. We optimize the RIS configuration for five distinct end-to-end communications performance indicators based on the considered, experimentally calibrated models. Then, we experimentally measure the wireless channel matrix for each optimized RIS configuration. We examine to what extent the gap in model accuracy translates into a gap in end-to-end performance. We also examine to what extent the models faithfully predict the achieved performances. 

This section constitutes the first experimental MNT-based performance evaluation for a large RIS. In Sec.~\ref{subsec_KPIs}, we define the five considered key performance indicators (KPIs). In Sec.~\ref{subsec_opti}, we describe our MC-aware discrete RIS optimization algorithm. In Sec.~\ref{subsec_exp_perf_results}, we present and discuss our experimental end-to-end performance evaluation.

\subsection{Key Performance Indicators (KPIs)}
\label{subsec_KPIs}

We consider two SISO KPIs and three MIMO KPIs. These five KPIs are defined as follows.

\textit{SISO KPI 1:} We aim to maximize the channel gain $|h_{ij}|^2$. This is a commonly considered KPI in theoretical papers on RIS~\cite{gradoni_EndtoEnd_2020,qian2021mutual,li2024beyond}. Albeit on the antenna level, this KPI directly correlates with numerous end-to-end SISO communications metrics (e.g., capacity, bit-error rate, throughput).

\textit{SISO KPI 2:} We assume that the $i$th TX communicates with the $i$th RX, such that the signals from the $i$th TX to the $j$th RX, where $j\neq i$, constitute undesired interferences. 
The rate achieved by the $i$th TX is hence $R_i=\mathrm{log}_2\left( 1 + P_\mathrm{T}|h_{ii}|^2 / (\sum_{j\neq i}  P_\mathrm{T}|h_{ij}|^2 + \sigma^2) \right)$, where we assume that all TX transmit with power $P_\mathrm{T}$; $\sigma$ quantifies the noise strength. This scenario represents a SISO interference channel. Our KPI is $R_i$, assuming $P_\mathrm{T}/\sigma^2=100$~dB.

\textit{MIMO KPI 1:} We aim to maximize the capacity $C$ at low SNR by dominant eigenmode transmission. In this regime, $C=\mathrm{log}_2\left( 1+ P_\mathrm{T}\|\mathbf{H}\|^2/\sigma^2 \right)$ so that maximizing $C$ is equivalent to maximizing the spectral norm of the channel matrix: $\|\mathbf{H}\|^2$. Hence, our KPI is $\|\mathbf{H}\|^2$.

\textit{MIMO KPI 2:} We assume the same scenario as in SISO KPI 2 but seek to maximize the sum rate as opposed to an individual rate. Our KPI is $R=\sum_i R_i$, assuming $P_\mathrm{T}/\sigma^2=100$~dB.

\textit{MIMO KPI 3:}  We aim to maximize the capacity $C$ at high SNR by multi-eigenmode transmission with uniform power allocation. In this regime,  $C=\mathrm{log}_2\left( \mathrm{det}\left( \mathbf{I}_{N_\mathrm{R}}+ P_\mathrm{T}\mathbf{H}\mathbf{H}^H/\sigma^2 \right)\right)$. Hence, our KPI is $C$, assuming $P_\mathrm{T}/\sigma^2=100$~dB.

Given the $4 \times 4$ MIMO channel in our experiment described in Sec.~\ref{sec_exp_validation} (see Fig.~\ref{fig2}), we can define 16 different versions of SISO KPI 1 and four different versions of SISO KPI 2. Thus, we separately optimize the RIS configuration for each version in turn, and we report the average of the optimized KPI over all versions in these two cases.

\subsection{MC-Aware Discrete RIS Optimization}
\label{subsec_opti}

Optimizing the RIS configuration based on a system model to maximize a given KPI is typically complicated. The problem is \textit{high-dimensional} for large-scale RIS; the problem is \textit{non-linear} due to MC governing the mapping from RIS configuration to channel, and the mapping from channel to KPI is usually also non-linear; the problem is \textit{discrete} because only a few loads are available at each RIS element (two for the RIS elements in our prototype); the problem is also almost always \textit{non-convex}. A closed-form solution only exists for a specific KPI under the assumption of an idealized fully-connected BD-RIS without any hardware constraints~\cite{nerini2024global}. Existing theoretical works on MC-aware RIS optimization assume that the reflection coefficients of the tunable lumped elements can take any value on or within the unit circle~\cite{gradoni_EndtoEnd_2020,qian2021mutual,abrardo2021mimo,ma2023ris,el2023optimization,li2024beyond,abrardo2024design,wijekoon2024phase,matteo_universal,nerini2024global,semmler2024performance,peng2025risnet}. Hence, these algorithms cannot be applied to our RIS prototype with 1-bit programmable elements. MNT-based discrete optimization has received little theoretical attention. Numerical and experimental works like~\cite{tapie2023systematic,del2019optimally} facing discrete constraints in RIS optimization opted for coordinate-descent algorithms. Our recent theoretical study in~\cite{cheima2025} suggests that coordinate descent indeed yields a good trade-off between performance and computational cost.

In light of these considerations, we use the same coordinate-descent algorithm to optimize our RIS based on the five considered models (MNT, MNTGFT, LFMNT, CASC, LR) for the five considered KPIs defined in Sec.~\ref{subsec_KPIs}. In each case, we define the objective function $\mathcal{O}$ to be minimized as the negative of the considered KPI. As initialization, we choose the configuration out of the $n_\mathrm{MNT}$ experimentally measured ones that yields the smallest objective function. Then, we loop over the RIS elements until the objective function has not decreased for $N_\mathrm{S}$ iterations. Algorithm~\ref{Alg_CoordDescent} summarizes our optimization procedure.

\begin{algorithm}
\footnotesize
\label{Alg_CoordDescent}
Evaluate the objective functions $\left\{\mathcal{O}\right\}_{\mathrm{init}}$ for the $n_\mathrm{MNT}$ experimentally measured RIS configurations $\left\{\mathbf{c}\right\}_{\mathrm{init}}$.\\
Select $\mathcal{O}_{\rm{curr}}$ as the minimum from $\left\{\mathcal{O}\right\}_{\mathrm{init}}$ and $\mathbf{c}_{\mathrm{curr}}$ from $\left\{\mathbf{c}\right\}_{\mathrm{init}}$ at the same index such that $\mathcal{O}_{\mathrm{curr}} = \mathcal{O}\left(\mathbf{c}_{\mathrm{curr}}\right) = \mathrm{min}\left(\left\{\mathcal{O}\right\}_{\mathrm{init}}\right)$.\\
$k \gets 0$, $i \gets 0$\\
\While{$k < N_\mathrm{S}$}{
    $i \gets i+1$\\
    \(j \gets \bigl((i-1) \bmod N_\mathrm{S}\bigr) + 1\)\\
    $\mathbf{c}_\mathrm{temp} \gets \mathbf{c}_\mathrm{curr}$ with $j$th entry flipped.\\
    $\mathcal{O}_{\rm temp} \gets \mathcal{O}\left(\mathbf{c}_\mathrm{temp}\right)$.\\
    \eIf{
    $\mathcal{O}_{\rm temp} < \mathcal{O}_{\rm curr}$
    } {
    $\mathbf{c}_\mathrm{curr} \gets \mathbf{c}_\mathrm{temp}$ \\
    $\mathcal{O}_\mathrm{curr} \gets \mathcal{O}_\mathrm{temp}$ \\
    $k \gets 0$
    } {$k \gets k+1$}
    }
\KwOut{$\mathbf{c}_{\rm curr}$ and $\mathcal{O}_\mathrm{curr}$.}
\caption{MC-Aware Binary RIS Optimization.}
\end{algorithm}

\subsection{Experimental Performance Evaluation Results}
\label{subsec_exp_perf_results}

\begin{table*}
\centering
\caption{Experimentally measured KPIs with RIS configurations optimized to maximize the KPIs based on the considered, experimentally calibrated models. The models' predictions for the maximized KPI values are indicated in gray font. RAND corresponds to the average of the measured KPI value across 300 random RIS configurations. }
\begin{tabular}{
    |p{2.1cm}||
    >{\centering\arraybackslash}p{0.75cm}||
    >{\centering\arraybackslash}p{0.75cm}|
    >{\centering\arraybackslash}p{0.75cm}||
    >{\centering\arraybackslash}p{0.75cm}|
    >{\centering\arraybackslash}p{0.75cm}||
    >{\centering\arraybackslash}p{0.75cm}| 
    >{\centering\arraybackslash}p{0.75cm}||
    >{\centering\arraybackslash}p{0.75cm}|
    >{\centering\arraybackslash}p{0.75cm}||
    >{\centering\arraybackslash}p{0.75cm}|
    >{\centering\arraybackslash}p{0.75cm}|
}
\hline
\multirow{2}{*}{KPI \textbackslash{} Method} & 
\multicolumn{1}{c||}{\textbf{RAND}} & 
\multicolumn{2}{c||}{\textbf{MNT}} & 
\multicolumn{2}{c||}{\textbf{MNTGFT}} & 
\multicolumn{2}{c||}{\textbf{LFMNT}} & 
\multicolumn{2}{c||}{\textbf{CASC}} & 
\multicolumn{2}{c|}{\textbf{LR}} \\
\cline{2-12}
& MEAS & MEAS & \textcolor{gray}{PRED} & MEAS & \textcolor{gray}{PRED} & MEAS & \textcolor{gray}{PRED} & MEAS & \textcolor{gray}{PRED} & MEAS & \textcolor{gray}{PRED} \\
\hline\hline
$|h_{ij}|^2 \! \times \! 10^3$ & 0.76 & 5.15 & \textcolor{gray}{5.16} & 5.16 & \textcolor{gray}{5.18} & 5.15 & \textcolor{gray}{5.61} & 4.38 & \textcolor{gray}{6.15} & 4.38 & \textcolor{gray}{5.51} \\
$R_i  $ [bits/s/Hz] & 0.59  & 5.67 & \textcolor{gray}{5.71} & 5.94 & \textcolor{gray}{6.09} & 5.67 & \textcolor{gray}{3.64} & 2.78 & \textcolor{gray}{7.22} & 2.28 & \textcolor{gray}{2.83} \\
$\|\mathbf{H}\|^2 \! \times \! 10^3$ & 8.40  & 15.70 & \textcolor{gray}{15.70} & 15.28 & \textcolor{gray}{15.26} & 15.70  & \textcolor{gray}{16.35} & 14.54 & \textcolor{gray}{17.52} & 13.91 & \textcolor{gray}{21.33} \\
$\sum_i R_i $ [bits/s/Hz] & 2.37   & 10.75 & \textcolor{gray}{10.71} & 8.91 & \textcolor{gray}{8.85} & 10.75 & \textcolor{gray}{6.80} & 5.31 & \textcolor{gray}{12.07} & 3.86 & \textcolor{gray}{5.81} \\
$C$ [bits/s/Hz] & 97.64  & 102.72 & \textcolor{gray}{102.72} & 102.76 & \textcolor{gray}{102.76} & 102.72 & \textcolor{gray}{102.72} & 102.05 & \textcolor{gray}{103.18} & 102.44 & \textcolor{gray}{104.39} \\
\hline
\end{tabular}
\label{table1}
\end{table*}

The KPIs achieved with the optimized RIS configurations are summarized in Table~\ref{table1}. Specifically, for each KPI we provide the value obtained on average over 300 random configurations (``RAND''), as well as with a configuration optimized based on each of the five considered models. We use the model parameters obtained in the calibrations with $p=1$ and $N_\mathrm{R}=N_\mathrm{T}=4$.
Both the model-based prediction of the KPI (``PRED'') and the corresponding experimentally measured KPI (``MEAS'') are provided.

For MNT and MNTGFT, we observe a very close agreement between prediction and measurement for all five KPIs. This observation not only reflects the high model accuracies already seen in Sec.~\ref{subsec_exp_param_estim} but  also testifies to the models' generalizability. Indeed, the models were calibrated based on measurements corresponding to random RIS configurations while the optimized RIS configurations that maximize a KPI may be distributional outliers. Accurate predictions of KPIs are very important for wireless practitioners to inform decisions on resource allocation; a simple example would be which wireless link(s) to enhance with a RIS in a multi-user scenario. In contrast, we see notable discrepancies between prediction and measurement for the three low-fidelity models which tend to significantly overestimate the achieved KPIs.

The achieved performance gains relative to the RAND benchmark vary significantly across the five considered KPIs. Based on our experimentally calibrated MNT model, we achieve a 6.8-fold and 9.6-fold improvement over RAND for SISO KPI 1 and SISO KPI 2, respectively. For MIMO KPI 2, the MNT-based improvement over RAND is 4.5-fold. For MIMO KPI 1, the MNT-based improvement over RAND is 87.0~\%. For MIMO KPI 3, the MNT-based improvement over RAND is only 5.2~\%. The small improvement for MIMO KPI 3 makes sense because the rich-scattering conditions combined with the assumed SNR of 100~dB are already highly ideal conditions on which it is hard to improve substantially by tuning the channel matrix with the RIS. Generally, it appears that the SISO KPIs depend more strongly on the RIS configuration and are thus more amenable to benefiting from RIS optimization.

The results with the MNTGFT model reveal no substantial performance benefits originating from the final optional global fine-tuning of the model parameters. In fact, surprisingly, we find that the achieved MIMO KPI 1 and MIMO KPI 2 are actually slightly inferior. However, the MNTGFT model does predict the RIS configurations optimized with the MNT model to be better, revealing that the simple coordinate-descent algorithm simply got stuck in local minima in these two cases. These are thus artefacts originating from inefficient model-based optimization as opposed to model inaccuracies. Meanwhile, the results of LFMNT perfectly coincide with those of MNT because the optimized RIS configuration was the same for all five KPIs. We observe thus no difference between the performances achieved with MNT, MNTGFT and LFMNT across five diverse KPIs. Nonetheless, only MNT and MNTGFT accurately predict the KPIs achieved based on them. 

The performances achieved with the MC-unaware CASC and LR models are still substantial but clearly inferior to those achieved with MNT, MNTGFT or LFMNT. Compared to RAND, CASC and LR still achieve roughly a 5.7-fold performance improvement for SISO KPI 1. CASC resp. LR achieve a 4.7-fold resp. 3.9-fold improvement for SISO KPI 2, a 73.2~\% resp. 65.7~\% improvement for MIMO KPI 1, a 2.2-fold resp. 1.6-fold improvement for MIMO KPI 2, and a 4.5~\% resp. 4.9~\% improvement for MIMO KPI 3. These substantial improvements are surprising given the low accuracies observed in Sec.~\ref{subsec_exp_param_estim} for the CASC and LR models. We hypothesize that the apparent robustness to model inaccuracies originates from the 1-bit constraint, in line with similar recent observations for small-scale RIS in~\cite{tapie2025beyond,del2025experimental}.

Altogether, models with limited or no MC awareness perform surprisingly well compared to MC-aware models in terms of the achieved KPIs. This observation questions whether the additional effort required to estimate the MC parameters is worthwhile. However, these lower-fidelity models predict the achieved KPIs rather inaccurately, which can be problematic for wireless practitioners facing resource allocation trade-offs.

\section{Discussion}
\label{sec_discussion}

Prior to this article, there were three important open questions from the experimental perspective on the role of MC awareness in RIS optimization:
\begin{enumerate}
    \item Do the benefits of MC awareness in RIS optimization found in theoretical studies manifest in experiments?
    \item Is it possible at all to experimentally acquire MC awareness for large RIS?
    \item Can the benefits outweigh the cost of experimentally acquired MC awareness?
\end{enumerate}
Our key findings are summarized in the following answers:
\begin{enumerate}
    \item The benefits in terms of achieved KPIs appear to be moderate, at least for few-bit-programmable RIS with half-wavelength inter-element spacing. However, the model-based predictions of the achieved KPIs are substantially more accurate with MC awareness.
    \item Yes.
    \item Possibly, e.g., in scenarios requiring accurate KPI predictions for resource allocation.
\end{enumerate}

Given recent enthusiasm for MC awareness in RIS optimization, our rather nuanced conclusions in 1) and 3) are possibly surprising. Yet, or maybe precisely for that reason, they constitute valuable takeaways for wireless practitioners, potentially saving them substantial calibration efforts dedicated to estimating the MC matrix. 

The quadratic scaling of the number of unknowns, and thus the number of required measurements, with the number of RIS elements is a fundamental property of the MNT model. As such, it is not possible nor our intention to fundamentally modify this scaling. Instead, we accommodate the consequences of this scaling: our segmentation ensures the feasibility of the required computations for each segment. Without segmentation, the computational requirements would grow quadratically with the number of RIS elements and inevitably overwhelm any given computational resource for a sufficiently large RIS.

Looking forward, we do not anticipate our method to be used in its current form once per coherence time in RIS-parametrized radio environments. Instead, we envision the following three use cases:

\begin{enumerate}
    \item \textit{Academic research:} Rigorous benchmarking of simplified lower-fidelity models requires well-controlled experiments with access to a high-fidelity ground-truth MC-aware model that our technique provides.
    \item \textit{One-off initialization:} In dynamic RIS-parametrized radio environments, many scattering objects are not dynamic. Thus, our technique can be applied once and the resulting parameters can subsequently serve as initialization for fine-tuning during each coherence time. It may furthermore be possible to only refine the most relevant parameters during each coherence time.
    \item \textit{Non-dynamic use cases:} Certain instances of massively parametrized wave systems are distinguished by the absence of uncontrolled perturbations. Examples include RIS-parametrized wireless networks-on-chip~\cite{tapie2023systematic}, dynamic metasurface antennas (DMAs), and wave-domain physical neural networks (PNNs). In these cases, it is sufficient to apply our technique once during a one-off system model calibration.
\end{enumerate}

\section{Conclusion}
\label{sec_conclusion}

To summarize, we reported the first experimental characterization of MC for a large-scale RIS prototype, and its subsequent use for MNT-model-based RIS optimization in diverse performance evaluations. To facilitate the required estimation of the 5867 MNT model parameters, we introduced and experimentally validated an ambiguity-aware segmentation of the estimation problem. Our systematic analysis revealed the importance of high-fidelity MC characterization to achieve high model accuracies. We further evidenced that larger MIMO systems enable higher model accuracies with fewer measurements, and that the model accuracy [dB] scales roughly linearly with the SNR [dB]. The substantial accuracy differences between models with full, partial or no MC awareness did \textit{not} translate into proportionate performance gain differences. Even MC-unaware models achieved strong performance improvements. Nonetheless, the important ability to predict the achieved performances based on the model did strongly correlate with the model accuracy. 

Looking forward, \textit{on the algorithmic side}, we encourage further investigations beyond the scope of this study to identify systematic guidelines for choosing the hyperparameters ($g$, $v$, $n_1$, $n_2$, $n_3$) of our proposed technique. Alternative solvers for the individual subproblems can also be explored.  Furthermore, identifying data-driven or prior-knowledge-based routes to determining the most and/or least influential entries of the MC matrix are of interest. Moreover, efficient updates of outdated model parameters will be relevant for dynamic use cases. Finally, it appears feasible to modify our technique to cope with purely non-coherent detection~\cite{sol2024experimentally,del2024virtual,del2024virtual2p0}, in order to alleviate the hardware requirements. Meanwhile, \textit{on the hardware side}, it will be interesting to determine whether the observed discrepancies between model accuracy and wireless performance generalize to RIS with continuously tunable and/or more closely spaced elements. Beyond RIS, we expect our technique to find applications for DMAs and PNNs.

\section*{Acknowledgment}
The author acknowledges I.~Ahmed, F. Boutet, and C. Guitton, who, under the author's supervision, previously built the RIS prototype for the work presented in~\cite{ahmed2025over}. The author further acknowledges stimulating discussions with L.~Le~Magoarou.

\bibliographystyle{IEEEtran}

\providecommand{\noopsort}[1]{}\providecommand{\singleletter}[1]{#1}%

\end{document}